\documentclass[10pt]{iopart}

\usepackage{iopams}

\usepackage{graphicx}
\usepackage[table]{xcolor}

\usepackage[numbers]{natbib}

\usepackage{dsfont}

\usepackage{bm}
\renewcommand{\vec}[1]{\bm{#1}}

\newcommand{\text}[1]{\mathrm{#1}}

\newcommand{\operatorname}[1]{\,\mathrm{#1}\,}

\newcommand{\bg}{\begin{eqnarray}}

\newcommand{\eg}{\end{eqnarray}}

\newcommand{\I}{i}

\newcommand{\D}{\text{d}}

 \newcommand{\bra}[1]{\left\langle{#1}\right|}
 \newcommand{\ket}[1]{\left|{#1}\right\rangle}
 \newcommand{\braket}[2]{\langle{#1}|{#2}\rangle}

\makeatletter
 \def\@makefnmark{\hbox{\@textsuperscript{\normalfont\@thefnmark}}}
\renewcommand\@makefntext[1]%
     {\noindent\makebox[0pt][r]{\textsuperscript{\@thefnmark}\,}#1}
\makeatother

\begin{document}

\title[]{Influence of pump coherence on the quantum properties of spontaneous parametric down-conversion}

 \author{Published in
    \href{https://doi.org/10.1088/1402-4896/aace12}
    {Physica Scripta \textbf{93}, 084001  (2018)}.}

\author{Enno Giese$^1$, Robert Fickler$^1$, Wuhong Zhang$^{1,2}$, Lixiang Chen$^{2}$ and Robert W. Boyd$^{1,3}$}

\address{$^1$Department of Physics, University of Ottawa, 25 Templeton Street, Ottawa, Ontario K1N 6N5, Canada.}
\address{$^2$Department of Physics, Xiamen University, Xiamen, 361005, China.}
\address{$^3$Institute of Optics, University of Rochester, Rochester, New York 14627, USA.}

\ead{egiese@uottawa.ca}

\begin{abstract}
The correlation properties of the pump field in spontaneous parametric down-conversion are crucial in determining the degree of entanglement of generated signal and idler photons. We find theoretically that continuous-variable entanglement of the transverse positions and momenta of these photons can be achieved only if the coherence of the pump beam is sufficiently high. The positions of signal and idler photons are found to be correlated, even for an incoherent pump. However, the momenta of the signal and idler photons are not anti-correlated, even though transverse momentum is conserved.
\end{abstract}
\ioptwocol

\setcounter{footnote}{0}

\section{Introduction}
Entanglement is not only a primary resource for quantum information processing, but also one of the fundamental concepts of physics.
Already in the early days of quantum mechanics, entanglement of continuous variables was proposed in the famous EPR paper~\cite{Einstein35} to discuss its implications on reality.

Nowadays, entangled photons are often generated by spontaneous parametric down-conversion (SPDC), the working horse in any quantum optics lab.
This nonlinear process is usually driven by a pump beam with transverse spatial coherence, whose profile not only determines the quality of entanglement~\cite{Howell04,Law04}, but also the transverse structure of the generated photons~\cite{Monken98,Chan07,Gomes09} and is used to perform foundational experiments~\cite{Menzel12,Menzel13} with implications on complementarity\footnote{
On the occasion of his 60$^\text{th}$ birthday and in light of his fascination for fundamental principles like complementarity and entanglement, we dedicate this article to Wolfgang P. Schleich.
\emph{Herzlichen Gl\"uckwunsch} and best wishes!
}.
Continuous-variable entanglement can be heuristically understood in terms simple arguments based for example on momentum conservation.
In this article we examine whether these fundamental arguments are still valid if we change the transverse coherence of the pump.

The significance of temporal coherence properties was investigated in~\cite{Burlakov01,Jha08,Kulkarni17}.
Even the transfer of transverse spatial coherence from the pump to the down-converted light was shown in~\cite{Jha10}, where the focus was on spatial two-qubit states and not primarily on high-dimensional entanglement.
In the present article, we hope to fill this gap and provide new insights in the role of coherence for the generation of entanglement.

\subsection{Correlations of two-dimensional distributions}
\label{subsec_Correlations}

Since we are interested in the continuous-variable entanglement of a photon pair created by SPDC, we first establish some concepts and terms in order to define entanglement.
A conditional probability distribution $P(v_s,v_i)$ may depend on two parameters $v_s$ and $v_i$, which will be later on assigned to the signal and idler photons, respectively.
We further introduce $v_\pm \equiv (v_s \pm v_i)/\sqrt{2}$, which corresponds to a coordinate system rotated by $\pi/2$, and the variances $\Delta v_\pm^2 $ along these axes.

\begin{figure}
\includegraphics[width=.5\columnwidth]{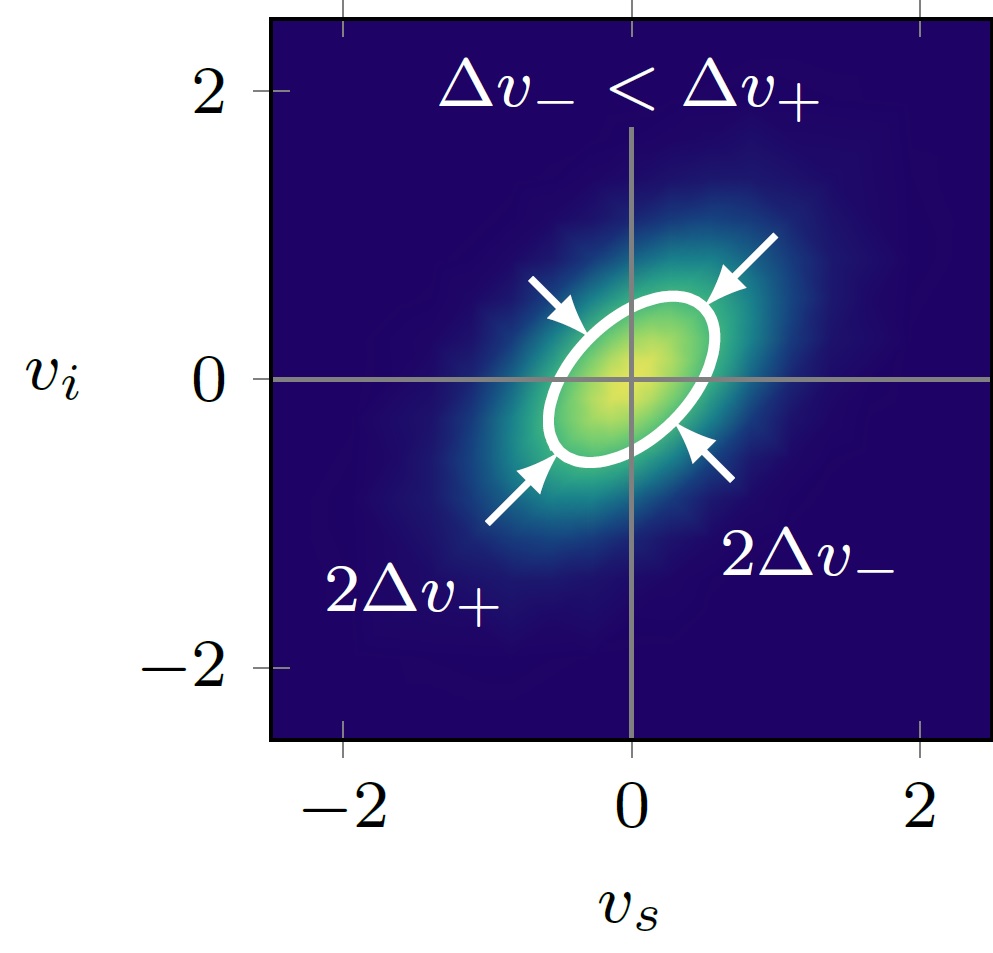}
\includegraphics[width=.5\columnwidth]{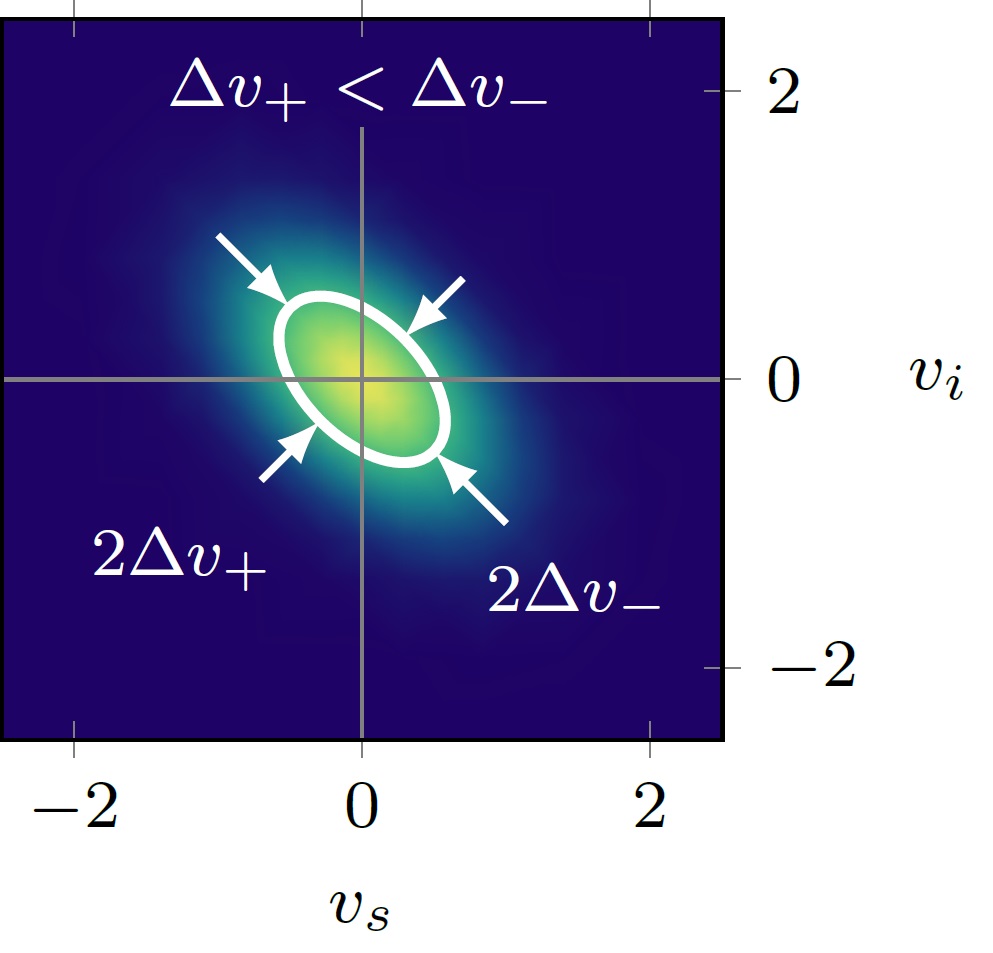}
\caption{Correlation of $v_s$ and $v_i$ (left), anti-correlation (right) and their connection to the width $\Delta v_\pm$. We show a density plot of the conditional probability distribution $P(v_s,v_i)$. The white contour corresponds to $1/\sqrt{\e}$ of the maximum.
	}
	\label{fig_correlations}
\end{figure}
We call the two parameters $v_s$ and $v_i$ \emph{correlated}, if the standard deviation of the anti-diagonal and the diagonal fulfill the relation $\Delta v_- < \Delta v_+ $, as depicted in figure~\ref{fig_correlations} on the left.
Knowing one of the two parameters, one can predict with a high probability that the other parameter is the same, i.e., $v_s\cong v_i$. 
Similarly, we call the two parameters $v_s$ and $v_i$ \emph{anti-correlated}, if the relation $\Delta v_+ < \Delta v_- $ is fulfilled, as depicted in figure~\ref{fig_correlations} on the right.
In this case, the two parameters most likely possess opposite values, i.e., $v_s\cong- v_i$. 
The two parameters are \emph{unccorrelated} if the probability distribution is separable, for example as in $P(v_s,v_i)=P_s(v_s)P_i(v_i)$, which means that knowing one parameter does not give any information about the other.
For Gaussian distributions this reflects itself in $\Delta v_+ = \Delta v_- $.
Correlations and anti-correlations exist for both classical and quantum mechanical probability distributions and do not necessarily have implications on the entanglement of a physical system.
However, entanglement can be revealed by simultaneous correlations of conjugate variables.

\subsection{Continuous-variable entanglement}
\label{subsuec_Continuous-variable}

We call two particles \emph{entangled} if their state cannot be written as a product of the state of each particle.
When we consider continuous-variable entanglement, in particular of momentum and position, the non-separabiltiy of an entangled state reflects itself in correlations.
However, such correlations may also exist for mixed states and the observation of correlations is not a sufficient entanglement witness.
One of the criteria for continuous-variable entanglement of momentum and position is the apparent violation of a Heisenberg-type relation~\cite{Reid89,Teh14}
\bg\label{eq_Heisenberg}
\Delta \mathcal{X} \Delta \mathcal{P}  \geq \hbar /2,
\eg
where $\Delta\mathcal{X}^2$ is the variance of a suitable linear combination of their positions of the sub-systems and $\Delta \mathcal{P}^2$ the variance of a linear combination of their momenta.
In the spirit of the EPR paradox~\cite{Einstein35}, a violation of this relation is not possible with classically correlated particles, since position and momentum are complementary variables.

\subsection{Entanglement in parametric down-conversion}
\label{subsec_Entanglement_PDC}

In the process of SPDC, one photon from a strong light field called the pump is converted by a nonlinear crystal into two photons called the signal and idler~\cite{Hong85}, see figure~\ref{fig_spdc}(a).
Multiple different types of entanglement of these photons have been demonstrated so far, such as polarization entanglement~\cite{Freedman72} or entanglement of orbital angular momentum~\cite{Mair01,Leach10}.
But also continuous variables entanglement has been observed, such as time and energy~\cite{Franson89,Kwiat93} or, more relevant for our discussion, in the transverse variables of signal and idler photons~\cite{Howell04,DAngelo04}.

Whether these photons are actually entangled depends crucially on the shape of the pump beam but also on the properties of the nonlinear crystal.
A usual heuristic argument for their entanglement is that a pump photon is converted into signal and idler at one particular position and hence their transverse positions $\vec{\varrho}_s$ and $\vec{\varrho}_i$ are correlated.
At the same time, the momentum of the pump photon is conserved and we therefore expect an anti-correlation of their transverse wave vectors $\vec{q}_s$ and $\vec{q}_i$.
The combination of position correlation and momentum anti-correlation gives, according to section~\ref{subsec_Correlations}, a small value of the product $ \Delta \vec{\varrho}_{-}\Delta \vec{q}_{+}$,
which may violate the condition for entanglement from \eref{eq_Heisenberg}.
Whereas the first quantity depends on phase matching, the second one is determined by the pump field.
Based on these arguments, one can even define a birth zone of the biphoton~\cite{Schneeloch16}.

For an infinite plane wave pump, i.e., a narrow angular distribution of the pump, the anti-correlations become perfect.
Different shapes of pump beams and their significance for the down-converted light have been investigated by several authors~\cite{Monken98,Chan07,Gomes09}.
In this article we focus on a different aspect of the pump field: its \emph{transverse coherence properties}.

\subsection{Outline of the article}
We start with section~\ref{sec_Spontaneous}, where we briefly review the process of SPDC, before we introduce the joint probability distribution of signal and idler photons in section~\ref{sec_Joint}.
The joint distribution consists of the product of the pump coherence function, which we specify in section~\ref{sec_Gaussian_Schell}, and a phase-matching function, derived in section~\ref{sec_Phase-matching}.
In section~\ref{sec_Entanglement} we investigate how the coherence of the pump changes the entanglement properties of the down-converted light, before we conclude in section~\ref{sec_Conclusions}. 
To keep this article self-contained, we introduce the Hamiltonian with a quantized pump field in  \ref{sec_Hamiltonian}, before we use it to calculate the biphoton density matrix in \ref{sec_Biphoton}.
We simplify our description by applying the Fresnel approximation in \ref{sec_Fresnel}.

\section{Spontaneous parametric down-conversion}
\label{sec_Spontaneous}
\begin{figure}
\includegraphics[scale=1]{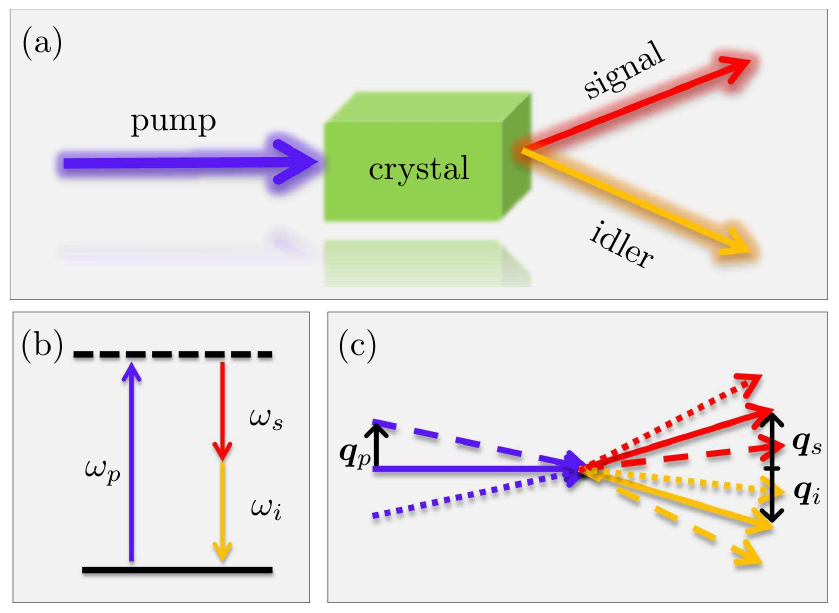}
\caption{Spontaneous parametric down-conversion of a pump beam generating signal and idler photons, schematically shown in part~(a).
A monochromatic pump implies energy conservation, shown in part~(b).
Transverse momentum is conserved, but the joint momentum distribution of signal and idler is determined by the distribution of the pump beam, shown in part~(c).
	}
	\label{fig_spdc}
\end{figure}

In the process of SPDC, a photon from a strong pump beam ($p$) is spontaneously converted by a nonlinear crystal into a photon pair consisting of signal ($s$) and idler ($i$) photons.
We depict the situation in figure~\ref{fig_spdc}(a) and we discuss the details of the quantum description in \ref{sec_Hamiltonian}.
Because we only consider the spontaneous regime and a monochromatic pump, energy conservation is fulfilled and we have $\omega_p = \omega_s+\omega_i$ for the frequencies of pump, signal and idler, depicted in figure~\ref{fig_spdc}(b).

Since the crystal is large in transverse direction compared to the beam, we also have transverse momentum conservation, i.e., $\vec{q}_p = \vec{q}_s + \vec{q}_i$, where $\vec{q}_j$ represents the transverse wave vector of the field $j$.
As we show in figure~\ref{fig_spdc}(c), a transverse momentum profile of the pump leads to a specific profile of the joint momentum distribution of signal and idler~\cite{Monken98}.
However, we show in \ref{sec_Biphoton} that it is \emph{not} the transverse momentum distribution of the pump that is relevant~\cite{Rubin96}, but the \emph{angular correlation function} $\tilde{\Gamma}$ of the pump~\cite{Marchand72,Jha10}.

In contrast to the transverse direction, the finite thickness and the shape of the crystal allows for a longitudinal wave vector mismatch $\Delta \kappa = \kappa_p-\kappa_s-\kappa_i$, which is reflected in its \emph{phase-matching function} $\tilde{\chi}$.

\section{Joint probability distribution}
\label{sec_Joint}

The description becomes particularly easy for degenerate SPDC:
Within the Fresnel approximation from \ref{sec_Fresnel} we use the density matrix established in \ref{sec_Biphoton} to derive the joint (conditional) probability distribution of transverse momenta of signal and idler photons. 
In particular, only the sum and the difference of the transverse momenta of signal and idler appear in the joint distribution function.
This fact naturally leads to a rotated coordinate system when we introduce the variables 
\bg
\vec{q}_\pm \equiv (\vec{q}_s \pm \vec{q_i})/\sqrt{2}.
\eg
In this case, we find with \eref{e_diagonal_elements}, which we normalize to unity, as well as with \eref{e_Delta_kappa_Fresnel} the expression
\bg\label{e_joint_mom_distr}
\tilde{P}(\vec{q}_s,\vec{q}_i)= \tilde{P}_\Gamma(\vec{q}_+)\tilde{P}_\chi(\vec{q}_-)
\eg
for the joint probability distribution of the transverse momenta~\cite{Jha10}.
To distinguish between distributions in position and momentum space, we introduce the tilde symbol for the latter. 
In~\eref{e_joint_mom_distr} we introduced the probability distribution along the diagonal
\bg
\tilde{P}_\Gamma(\vec{q}_+) =  \tilde{\Gamma}(\sqrt{2}\vec{q}_+,\sqrt{2}\vec{q}_+)
\eg
and the probability distribution along the anti-diagonal
\bg\label{e_tilde_P_chi}
\tilde{P}_\chi(\vec{q}_-) = \left| \tilde{\chi} \left(  \vec{q}^2_-/k_p \right)\right|^2
\eg
in $(\vec{q_s},\vec{q}_i)$-space, which we assume both to be normalized.
According to \eref{e_diagonal_elements}, $\tilde{P}_\Gamma(\vec{q}_+)$ is determined by the angular correlation function of the pump $\tilde{\Gamma}$, and $\tilde{P}_\chi(\vec{q}_-)$ depends on the phase-matching function $\tilde{\chi}$.
Following our discussion from section~\ref{subsec_Correlations}, the ratio of their widths determines whether the momenta are correlated, anti-correlated, or uncorrelated.

Through use of the density matrix~\eref{e_density_matrix}, it is possible to transform the momentum distribution into position space, which leads to the expression
\bg
P(\vec{\varrho}_s,\vec{\varrho}_i)= P_\Gamma(\vec{\varrho}_+)P_\chi(\vec{\varrho}_-),
\eg
where we have
\bg
P_\Gamma(\vec{\varrho}_+) =  \Gamma(\vec{\varrho}_+/\sqrt{2},\vec{\varrho}_+/\sqrt{2})
\eg
that depends on the \emph{position correlation function} $\Gamma$ of the pump beam and
\bg\label{e_P_chi}
P_\chi(\vec{\varrho}_-) =\left|\frac{1}{(2\pi)^2}\int\D^2q_-\, \e^{-\I \vec{q}_-\vec{\varrho}_-} \tilde{\chi} \left(  \vec{q}^2_-/k_p \right)\right|^2,
\eg
which is the Fourier transform of the phase-matching function.
Here, we also defined a rotated coordinate system
\bg
\vec{\varrho}_\pm \equiv (\vec{\varrho}_s \pm \vec{\varrho}_i)/\sqrt{2}
\eg
with an analogous interpretation in the $(\vec{\varrho}_s,\vec{\varrho}_i)$-space.

Both $P(\vec{\varrho}_s,\vec{\varrho}_i)$ and $\tilde{P}(\vec{q}_s,\vec{q}_i)$ are separable in the $\vec{\varrho}_\pm$ and $\vec{q}_\pm$ coordinates.
That means that both distribution functions are in general not separable in a distribution that can be assigned to the signal photon and a distribution that can be assigned to the idler photon.
However, due to their separability in $\vec{\varrho}_\pm$ in the near field as well as in $\vec{q}_\pm$ in the far field, it is easy to obtain the variances $\Delta \vec{\varrho}_{\pm,j}^2$ and $\Delta \vec{q}_{\pm,j}^2$, where $j$ is one particular transverse component:
They correspond directly to the variances of the respective probability functions $P_{\Gamma,\chi}$ and $\tilde{P}_{\Gamma,\chi}$ along the diagonal and anti-diagonal.
This fact will become important when we discuss the entanglement criteria in section~\ref{sec_Entanglement}.
In the next two sections we discuss a specific model for the correlation function of the pump beam and the phase-matching function of a bulk crystal.

\section{Gaussian Schell-model pump beam}
\label{sec_Gaussian_Schell}
Since the diagonal (and by that the correlations) of the joint distribution in both position and momentum space is determined by the angular correlation function of the pump, we specify a model for the pump in this section.
Let us assume that the pump beam has the form of a monochromatic Gaussian Schell-model beam~\cite{Mandel95}.
This rather general beam allows us to look at the effect of the transversal coherence of the pump, of course only within the treatment of the paraxial approximation.

The two-point correlation function of a Gaussian Schell-model beam reads
\bg\label{e_GS_beam}
\Gamma(\vec{\varrho}_1,\vec{\varrho}_1)  \sim  \exp \left[- \frac{\vec{\varrho}_1^2+\vec{\varrho}_2^2}{4 w^2} - \frac{(\vec{\varrho}_1-\vec{\varrho}_2)^2}{2 \ell_c^2} - \I \frac{\vec{\varrho}_1^2-\vec{\varrho}_2^2}{2 \mathcal{R}^2} \right]\nonumber \\
\eg
and depends on the width $w$ of the beam, its transverse coherence length $\ell_c$, and on the radius of curvature $R$ and the wave vector $k_p$ of the pump through $\mathcal{R}^2 = R/k_p$.
Note that for our treatment we assume these quantities to be constant upon propagation within the crystal.

With~\eref{e_GS_beam}, the probability distribution reduces to
\bg\label{e_P_+_rho}
P_\Gamma(\vec{\varrho}_+)\sim \exp\left[- \vec{\varrho}^2_+/(4 w^2)\right],
\eg
which is independent of its transversal coherence length $\ell_c$ and solely depends on the width $w$ of the beam.
From \eref{e_GS_beam} we can also calculate the momentum correlation function.
Since we are only interested in the diagonal elements, the transformation can be easily performed, e.g., by using a Wigner representation~\cite{Schleich01}, and we find
\bg\label{e_P_+_q}
\tilde{P}_\Gamma(\vec{q}_+) \sim \exp\left[- \frac{4\ell_c^2 w^2\vec{q}^2_+}{4 w^2+ \ell_c^2 (1+ 4 w^4/\mathcal{R}^4)}\right].
\eg
This distribution depends on all quantities, the width of the pump beam, the coherence length as well as the radius of curvature.

Since the probability distribution functions \eref{e_P_+_rho} and \eref{e_P_+_q} are of Gaussian form, the variances can be easily obtained and are the same for both transverse directions $j$.
We find from \eref{e_P_+_rho} that
\bg\label{e_Delta_rho_+}
\Delta \vec{\varrho}^2_{+,j} = 2 w^2
\eg
for the variance of the position correlations and from \eref{e_P_+_q}
\bg\label{e_Delta_q_+}
\Delta \vec{q}^2_{+,j} = \frac{4 w^2+ \ell_c^2 \left(1+ 4 \frac{w^4}{\mathcal{R}^4}\right)}{8\ell_c^2 w^2} = \frac{1+ 4 \left(\frac{w^4}{\mathcal{R}^4}+\frac{w^2}{\ell_c^2}\right)}{8 w^2}
\eg
for the variance of the momentum correlations.
In the last step we rewrote the variance to show that it scales with the inverse of the variance in position if the beam has no curvature and is coherent.
However, a radius of curvature as well as a finite coherence length both contribute similarly and both increase the width of the momentum distribution.

Before we turn to the phase-matching function, we discuss the two limiting cases of a fully coherent and incoherent pump beam.
First, we take the limit $w\ll \ell_c$, that is, of a very large coherence length and find
\bg
\Delta \vec{q}^2_{+,j}\cong (1+ 4 w^4/\mathcal{R}^4)/(8 w^2),
\eg
which is the usual result.

However, in the opposite case where $\ell_c\ll w$, i.e., the beam is incoherent, we have
\bg
\Delta \vec{q}^2_{+,j} \cong 1/(2\ell_c^2 )
\eg
and see that it is solely determined by the transverse coherence length.
Hence, in case of an incoherent beam, we find that this variance increases rapidly.

\section{Phase-matching function}
\label{sec_Phase-matching}
The anti-diagonal (and by that the anti-correlations) of the joint distribution in both position and momentum space is determined by the phase-matching function of the crystal.
It may be obtained by the Fourier transformation of the spatial profile $\chi^{(2)}(z)$ of the nonlinearity along the longitudinal direction $z$, namely
\bg\label{e_phase-matching_def}
\tilde{\chi}(\Delta \kappa) \sim \int \D z\, \e^{\I \Delta \kappa z} \chi^{(2)}(z).
\eg
The modulus square of the phase-matching function gives directly the joint momentum distribution for the difference of the momenta, as can be seen from \eref{e_tilde_P_chi}.
The inverse Fourier transform in \emph{transversal} direction leads to the distribution of the difference of the positions, following \eref{e_P_chi}.
With this insight it is easy to understand the behaviour of different types of nonlinear crystals.
Even though we discuss a single homogeneous crystal in the following, we emphasize that~\eref{e_phase-matching_def} is a general expression and both quasi-phase matching~\cite{Boyd08} as well as nonlinear interferometers in the spontaneous regime~\cite{Klyshko93} can be easily understood in this formulation.

Let us assume that we have a uniform crystal of length $L$ with its front face positioned at $z_0$.
In this case, the nonlinearity $\chi^{(2)}(z)$ can be written as a constant that begins at $z_0$ and ends at $z_0+L$, namely
\bg\label{e_boxcar}
\chi^{(2)}(z) \sim \theta(z-z_0)-\theta(z-z_0-L),
\eg
where $\theta(z)= 1 $ for $z\geq 0$ and 0 otherwise.
The Fourier transform of this boxcar function is proportional to the cardinal sine function $\operatorname{sinc}(x) \equiv \sin (x)/x$, and we find 
\bg\label{e_chi_sinc}
\tilde{\chi}(\Delta \kappa) \sim L \exp[\I \Delta \kappa (z_0-L/2)] \operatorname{sinc} (\Delta \kappa L/2),
\eg
where we included the proportionality to $L$ to demonstrate that the coincidence rate scales with $L^2$.
In the following we drop this factor for simplicity.
The phase-matching function may have a phase that depends on $\Delta \kappa$, but also on $z_0$ and $L$.
Even though this phase does not change the behavior of the momentum correlations $\tilde{P}_\chi$, it has an effect on the position correlations $P_\chi$, since the phase is relevant for the inverse Fourier transform.
If $z_0=L/2$, which means that the crystal is centred around the origin, the phase appearing in $\tilde{\chi}$ vanishes.
In this case and within the Fresnel approximation, it is possible to perform the transformation~\eref{e_P_chi} analytically and to obtain the probability distribution function
\bg
P_\chi(\vec{\varrho}_-) \sim \left[\frac{\pi}{2} - \operatorname{Si}\left( \frac{k_p\vec{\varrho}_-^2}{2L}\right)\right]^2,
\eg
where $\operatorname{Si}(x) = \int_0^x\D y\, \sin(y)/y $ is the sine integral.
However, we assume that the origin is at the back face of the crystal and choose $z_0=L$.
In this case, the analytical Fourier transformation cannot be performed as easily.

\paragraph{Gaussian approximation}
For an intuitive understanding of the influence of phase matching on the correlations of signal and idler photons, a Gaussian approximation of the phase-matching function is routinely applied~\cite{Walborn10}, so that
\bg\label{e_chi_Gauss}
\tilde{\chi} (\Delta \kappa) \sim \exp[\I \vec{q}_-^2L/(2k_p)]\exp\left[- \alpha \vec{q}_-^2 L/(2 k_p)\right] ,
\eg  
which leads to
\bg
\tilde{P}_\chi(\vec{q}_-) \sim \exp\left[- \alpha \vec{q}_-^2 L/k_p\right] 
\eg 
with a variance
\bg\label{e_Delta_q_-}
\Delta \vec{q}_{-,j}^2 = k_p/(2\alpha L).
\eg
For $\alpha = 0.455$, which is a common choice~\cite{Walborn10}, both \eref{e_chi_sinc} and \eref{e_chi_Gauss} coincide where they both take their $1/\e$ value.

To obtain the position distribution, we Fourier transform \eref{e_chi_Gauss} and find
\bg
P_\chi(\vec{\varrho}_-) \sim \exp\left[- \frac{k_p \alpha }{L (1+\alpha^2)}\vec{\varrho}_-^2\right],
\eg
which has a variance
\bg \label{e_Delta_varrho_-}
\Delta \vec{\varrho}_{-,j}^2 = L (\alpha^{-1}+ \alpha)/(2k_p).
\eg

\section{Entanglement and pump coherence}
\label{sec_Entanglement}
In most experiments generating entanglement, the positions of signal and idler are correlated and their momenta anti-correlated, which we heuristically explained in section~\ref{subsec_Entanglement_PDC} basend on a birth zone of the photon~\cite{Schneeloch16} and momentum conservation.
We first discuss the opposite situation that is not often experimentally investigated, namely that there are position anti-correlations and momentum correlations.

With the help of \eref{e_Delta_rho_+} and \eref{e_Delta_q_-} we find that the product 
\bg\label{e_ent1}
\Delta \vec{\varrho}_{+,j}^2 \Delta \vec{q}_{-,j}^2 = k_p w^2 /(\alpha L)
\eg
depends on both the pump waist and the crystal length.
At least theoretically there is the possibility that the product is small by using a small pump waist $w$ or a long crystal.
However, for a tightly focused pump, the Fresnel approximation is not valid and other challenges arise.
SPDC from a highly focused pump has for example be observed in~\cite{DiLorenzoPires11}.
However, a few works are considering this case~\cite{Yun12,Zhong15,CalderonLosada16}, which is actually in contrast to the heuristic arguments for entanglement that we gave in section~\ref{subsec_Entanglement_PDC}.
For an extremely narrow pump beam or long crystal $w \ll L$ within the paraxial approximation it is possible to show entanglement~\cite{Zhong15} with $\Delta \vec{\varrho}_{+,j}\Delta \vec{q}_{-,j}< 1/2$.

\begin{figure}
\includegraphics[scale=1]{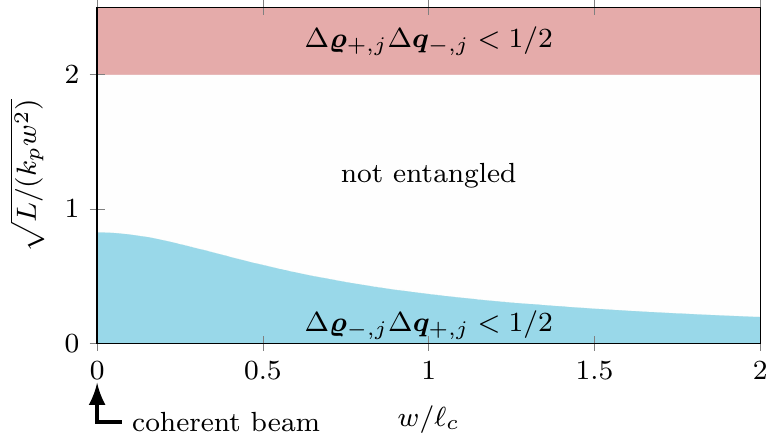}
\caption{Parametric plot of $\Delta \vec{\varrho}_{+,j} \Delta \vec{q}_{-,j}<1/2$ and $\Delta \vec{\varrho}_{-,j} \Delta \vec{q}_{+,j}< 1/2$ depending on $w/\ell_c$ and $\sqrt{L/(k_pw^2)}$.
In the colored areas of parameter space, one of the two conditions is fulfilled and we have entanglement, in the white area neither condition is fulfilled and no entanglement is present.
From left to right, the coherence of the pump decreases.
We chose $w\ll \mathcal{R}$.
In the blue shaded area, there is position correlation and momentum anti-correlation. In contrast, in the red shaded area we obtain the unusual situation of anti-correlation in position and correlation in momentum.
}
	\label{fig_entanglement}
\end{figure}

To see how the pump coherence affects entanglement, we could model the partially coherent beam as a mixture of many small coherence areas.
Therefore, a misleading argument is to think that the coherence length $\ell_c$ plays the role of the pump waist.
Since $\ell_c \ll L$, one would always expect very strong position anti-correlation and momentum correlation, thus leading to entanglement.
However, we show from \eref{e_ent1} that the coherence length is irrelevant for the position anti- and momentum correlations.
Therefore, the arguments given above are not correct.
The product $\Delta \vec{\varrho}^2_{+,j}\Delta \vec{q}^2_{-,j}$ is independent of the pump coherence, demonstrated in figure~\ref{fig_entanglement}.

For the opposite case (position correlation and momentum anti-correlation), we find from~\eref{e_Delta_varrho_-} and from \eref{e_Delta_q_+} that
\bg\label{e_ent2}
\Delta \vec{\varrho}_{-,j}^2 \Delta \vec{q}_{+,j}^2 = \frac{1 + \alpha ^{-2}}{16}\left(1+ \frac{4 w^4}{\mathcal{R}^4} + \frac{4 w^2}{\ell_c^2}\right) \frac{\alpha L}{k_p w^2},
\eg
which is not the inverse of~\eref{e_ent1} as one would naively expect from Fourier transform arguments.
At the same time, that means that for $\Delta \vec{\varrho}_{+,j}\Delta \vec{q}_{-,j}> 1/2$ we do not automatically have position correlation and momentum anti-correlation.
This is due to the phase factor of the phase-matching function and due to the finite coherence length $\ell_c$ and radius of curvature $\mathcal{R}$.
If \eref{e_chi_Gauss} had no phase factor, the factor $1+\alpha^{-2}$ would be replaced by unity and for $w\ll \mathcal{R},\ell_c  $ there is always some type of entanglement except for $\Delta \vec{\varrho}_{-,j} \Delta \vec{q}_{+,j}=\Delta \vec{\varrho}_{+,j} \Delta \vec{q}_{-,j}=1/2$.

We plot both possible conditions for entanglement $\Delta \vec{\varrho}_{+,j} \Delta \vec{q}_{-,j}<1/2$ and $\Delta \vec{\varrho}_{-,j} \Delta \vec{q}_{+,j}< 1/2$ in figure~\ref{fig_entanglement} as an area plot of the dimensionless parameters $w/\ell_c$ and $\sqrt{L/(k_pw^2)}$ under the assumption that $w\ll \mathcal{R}$.
The white area in parameter space shows conditions where we have no entanglement.
In the red shaded area the first condition is fulfilled and we have entanglement with position anti-correlation and momentum correlation.
We see that this condition is independent of the coherence length.

In contrast to that, the blue shaded area depicts the parameter regime where the second condition is fulfilled and we have entanglement with position correlation and momentum anti-correlation.
We see that the requirements for entanglement on $\sqrt{L/(k_pw^2)}$ get stronger the more we reduce the transverse coherence of the beam.

Indeed, the relation \eref{e_Delta_q_+} demonstrates that the momenta are not anti-correlated, if the coherence length or the radius of curvature is too small.
Therefore, entanglement eventually gets destroyed by decreasing coherence.

\begin{figure}
\includegraphics[scale=1]{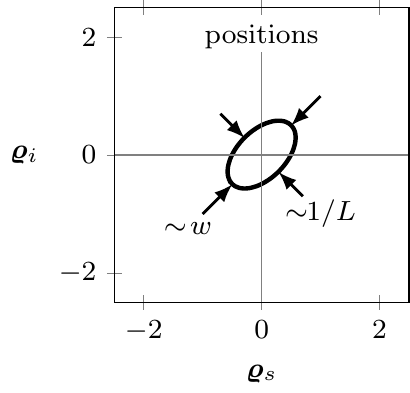}
\includegraphics[scale=1]{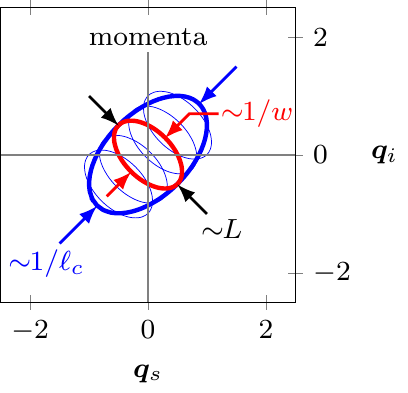}
\caption{Contour plot of the joint position (left) and momentum distribution (right) in arbitrary units.
Whereas there exist position correlations independently of the coherence of the pump, the momentum anti-correlations of a coherent pump (red contour) vanish for decreasing coherence (blue thick contour), which can be understood as a mixture of different coherent areas (blue thin contours).
For simplicity, we chose $w\ll \mathcal{R}$.}
	\label{fig_correlations2}
\end{figure}

We show in figure~\ref{fig_correlations2} contour plots to illustrate the difference between coherent and incoherent pump beams. 
On the left, the joint position distribution shows correlations independent of the coherence length.
On the right, we depict the joint momentum distribution for different input beams.
For a coherent beam, the momenta are anti-correlated (red contour).
However, for an incoherent beam (blue thick contour), we see that the width of the anti-diagonal distribution remains unchanged, but the width of the diagonal increases significantly, which is why the entanglement vanishes.
We still see position correlations, but the momenta are not anti-correlated but become either uncorrelated or even correlated.
One can understand the effect heuristically by assuming that there is an incoherent superposition of different coherence areas, each with a slightly different momentum distribution.
For each coherence area, momentum conservation is fulfilled (blue thin contours), but there they are shifted due to different pump momenta.
Since it is an incoherent mixture of these shifted distributions, the resulting distribution corresponds to the blue contour.

\section{Conclusions}

\label{sec_Conclusions}

We have shown that not only the pump profile but also its coherence is transferred to signal and idler fields in the process of SPDC~\cite{Jha10}.
In fact, it is not the pump profile that is relevant for the joint transverse distribution of signal and idler, but its angular correlation function.
For a Gaussian Schell-model beam, we observe that there are position correlations independent of the coherence of the beam.
The positions of signal and idler are always correlated, and this correlation depends only on the transverse spatial profile of the pump inside the crystal.
Even though there is transverse momentum conservation, this does not result in anti-correlated momenta of signal and idler photons.
The reason is that for an incoherent beam there is no narrow and coherent momentum distribution, just a mixture of them.
This mixture leads to a drastic change in the joint distribution and anti-correlations vanish.

Conventionally, SPDC is used to generate entanglement with position correlations and momentum anti-correlations (with a short crystal and wide beam).
Position correlations are therefore naturally present, but the  anti-correlations in momentum---necessary to observe the conventional type of entanglement---crucially depend on the coherence of the pump.
Hence, with a partially coherent beam it becomes increasingly harder to fulfil the requirements for entanglement.
For an incoherent pump beam, singal and ider photons are not entangled in their transverse continuous variables.
In the future, we plan to demonstrate this effect experimentally with truly transverse incoherent light.

%%% Acknowledgments
\ack{We thank Armin Hochrainer for valuable comments and discussions.
EG, RF, and RWB are thankful for the support by the Canada First Research Excellence Fund award on Transformative Quantum Technologies and by the Natural Sciences and Engineering Council of Canada (NSERC).
RF acknowledges the financial support of the Banting postdoctoral fellowship of the NSERC and WZ is thankful for the financial support of the China Scholarship Council (CSC).}

\appendix 
\section{Hamiltonian}
\label{sec_Hamiltonian}
To describe the process of SPDC, we use an heuristic Hamiltonian where the three fields of pump ($p$), signal ($s$) and idler ($i$) are coupled by a $\chi^{(2)}$-nonlinearity, shown in figure~\ref{fig_spdc}(a).
We perform an expansion into plane waves with wave vector $\vec{k}_j^\text{T} = (\vec{q}_j^\text{T}, \kappa_j)$, where $\vec{q}_j$ is the transversal and $\kappa_j$ the longitudinal wave vector of the field $j=p,s,i$.
Moreover, the superscript $\text{T}$ denotes the transpose.
These field modes are quantized and we introduce the creation and annihilation operators $\hat{a}_j^\dagger(\vec{k}_j)$ and $\hat{a}_j(\vec{k}_j)$ that fulfil the Bosonic commutation relation.

Within this expansion, it is possible to apply the rotating wave approximation.
Assuming a monochromatic pump gives directly the condition of energy conservation, i.e. $\omega_p = \omega_s+\omega_i$ shown in figure~\ref{fig_spdc}(b), where the frequencies $\omega_j \equiv c k_j/n_j$ with the modulus $k_j\equiv |\vec{k}_j|$ of the wave vector and the index of refraction $n_j$ of that particular field.
When we additionally assume that only one frequency for each field, namely $\omega_s$ and $\omega_i$, is being detected (for example by introducing filters), the operators $\hat{a}_j(\vec{k}_j)$ depend effectively only on the transverse momenta $\vec{q}_j$.
We therefore suppress from now on the dependence of the photon operators on the longitudinal wave vector $\kappa_j$.

If the nonlinear medium is infinite in the transversal direction, we find transverse momentum conservation $\vec{q}_p = \vec{q}_s+\vec{q}_i$, as shown in figure~\ref{fig_spdc}(c).
However, since the medium is finite in longitudinal direction, we do not have exact momentum conservation and phase mismatches $\Delta \kappa \equiv \kappa_p - \kappa_s-\kappa_i$ are allowed.
This fact is reflected in the phase-matching function $\tilde{\chi}(\Delta \kappa)$, which is the Fourier transform of the spatial shape of the $\chi^{(2)}$-nonlinearity in longitudinal direction.
We define it explicitly in \eref{e_phase-matching_def} in the main body of the text.

Neglecting walk-off effects and different polarizations, the effective interaction Hamiltonian
\bg
\hat{H} \sim \int\D^4q\, \tilde{\chi}(\Delta \kappa) \hat{a}_p(\vec{q}_s+\vec{q}_s)\hat{a}^\dagger_s(\vec{q}_s)\hat{a}^\dagger_s(\vec{q}_s) + \text{h.c.}
\eg
describes the nonlinear interaction.
The abbreviation $\text{h.c.}$ denotes the Hermitian conjugate.
Note that we integrate over both the transverse momentum of the signal and the idler photon.

\section{Biphoton density matrix}
\label{sec_Biphoton}
In the present appendix we introduce the biphoton density matrix that arises from the nonlinear interaction.
Since we are focussing on the spontaneous process, we assume that both the signal and the idler fields are initially in the vacuum state and uncorrelated to the pump field so that the initial density matrix reads 
\bg
\hat{\rho}_\text{ini} = \hat{\rho}_p \otimes \ket{0}_s\!\bra{0}\otimes \ket{0}_i\!\bra{0},
\eg
where $\hat{\rho}_p$ denotes the initial density matrix of the pump.

Furthermore, we can neglect higher-order processes in SPDC.
Thus, we may apply perturbation theory and the time-evolution operator $\mathds{1}-\I \hat{H}t/\hbar$.
We find this form only because we assumed a monochromatic pump, otherwise we would have to integrate in time over the pump spectrum.
However, the results would not change qualitatively and we therefore refrain from discussing it in more detail.
The time-evolved density matrix takes the form
\bg
\hat{\rho} = \hat{\rho}_\text{ini} - \frac{\I t}{\hbar}\left(\hat{H}\hat{\rho}_\text{ini} - \hat{\rho}_\text{ini} \hat{H}\right)+ \frac{t^2}{\hbar ^2}\hat{H}\hat{\rho}_\text{ini} \hat{H}.
\eg
Since we only detect the photons in the signal and idler fields and postselect on coincidence events, the first two terms do not contribute.
Therefore, we introduce the biphoton density matrix
\bg
\hat{\rho}_\text{bi}\sim \text{Tr}_p\left[\hat{H}\hat{\rho}_\text{ini} \hat{H}  \right],
\eg  
where we trace is over the pump field.
Because at most one signal and one idler photon are created in SPDC, we introduce the notation $\ket{\vec{q}_j}\equiv \hat{a}^\dagger_j(\vec{q}_j)\ket{0}_j $, which can be interpreted as a momentum eigenstate of the created photon with $\braket{\vec{q}_j'}{\vec{q}_j}=\delta(\vec{q}_j'-\vec{q}_j)$.
Therefore, we arrive at the density matrix
\bg\label{e_density_matrix}
\hat{\rho}_\text{bi}\sim \int\D^4q&\int\D^4q'\,\tilde{\rho}(\vec{q}_s, \vec{q}_i ; \vec{q}'_s,\vec{q}'_i) \ket{\vec{q}_s,\vec{q}_i}\bra{\vec{q}_s',\vec{q}_i'}
\eg
in momentum representation, where the density matrix elements
\bg\label{e_matrix_elements}
\tilde{\rho}(\vec{q}_s, \vec{q}_i ; \vec{q}'_s,\vec{q}'_i)\sim &\text{Tr}_p\left[\hat{a}_p(\vec{q}_s+\vec{q}_i)\hat{\rho}_p  \hat{a}_p^\dagger(\vec{q}_s'+\vec{q}_i') \right]\nonumber\\
&\times \tilde{\chi}(\Delta \kappa)\tilde{\chi}^*(\Delta \kappa') 
\eg
consist of a product of the \emph{transverse momentum correlation function}~\cite{Marchand72} of the pump field
\bg
\tilde{\Gamma}(\vec{q}_p,\vec{q}_p')\equiv \text{Tr}_p\left[\hat{a}_p(\vec{q}_p)\hat{\rho}_p  \hat{a}_p^\dagger(\vec{q}_p') \right]
\eg
and the phase-matching function of the crystal $\tilde{\chi}$.
The argument of the phase-matching function is the longitudinal phase mismatch between pump, signal and idler field and takes the form 
\bg\label{e_mismatch}
\Delta \kappa =  \kappa_p(\vec{q}_s+\vec{q}_i)- \kappa_s(\vec{q}_s)- \kappa_s(\vec{q}_s),
\eg
where we have applied the condition of transverse momentum conservation.
In analogy, $\Delta \kappa'$ is of the same form, only with $\vec{q}_j$ replaced by $\vec{q}_j'$

With the density matrix \eref{e_density_matrix} the joint probability distribution function can be defined as the diagonal elements of $\hat{\rho}_\text{bi}$, i.e.
\bg\label{e_diagonal_elements}
\tilde{\rho}(\vec{q}_s, \vec{q}_i ; \vec{q}_s,\vec{q}_i) \sim \left|\tilde{\chi}(\Delta \kappa)\right|^2 \tilde{\Gamma}(\vec{q}_s+\vec{q}_i,\vec{q}_s+\vec{q}_i).
\eg
This is a generalization of the conventional description, where usually the correlation function is replaced by the pump profile~\cite{Rubin96}.
A similar expression of the density matrix has been provided in~\cite{Jha10}, but without its explicit derivation.
If we were interested in the frequency spectrum rather than the angular spectrum, the temporal coherence of the pump would play the role of the angular correlation function~\cite{Burlakov01,Jha08,Kulkarni17}.

\section{Fresnel approximation}
\label{sec_Fresnel}
To simplify expression~\eref{e_mismatch}, we introduce the Fresnel approximation~\cite{Walborn10}.
We expand the longitudinal wave vector $\kappa_j = \sqrt{k_j^2 - \vec{q}_j^2}$ for $|\vec{q}_j|\ll k_j$ and arrive at  
\bg
\kappa_j  \cong k_j - \vec{q}_j^2/(2k_j)+ \cdots.
\eg
The approximation is valid only in the paraxial regime.
With this expression, we find
\bg
\Delta \kappa \cong k_p-k_s-k_i - \frac{(\vec{q}_s+\vec{q}_i)^2}{2 k_p}+\frac{\vec{q}_s^2}{2k_s}+\frac{\vec{q}_i^2}{2k_i}.
\eg
We further assume that together with energy conservation the fields fulfil $k_p = k_s+k_i$.
In this case, we find the simple expression
\bg\label{e_Delta_kappa_Fresnel}
\Delta \kappa \cong (\beta \vec{q}_s- \beta^{-1}\vec{q}_i)^2/(2 k_p),
\eg
where we defined the parameter $\beta^2 = k_i/k_s$ that is unity in the degenerate case.
Hence, the phase-matching function $|\tilde{\chi}(\Delta\kappa)|^2$ in~\eref{e_diagonal_elements} only depends on $\beta\vec{q}_s - \beta^{-1}\vec{q}_i$, i.e., on the weighted difference of the two transversal wave vectors.
We emphasize at this point that the paraxial approximation is of course not valid for any type of radiation, and for transverse incoherent light one has to make sure that one has a beam-like behavior.
Gaussian Schell-model beams might be one class of beams where the paraxial approximation is valid.
However, a more sophisticated treatment can be performed by taking the different dispersion relations of the three fields inside the crystal into account.
For simplicity and to obtain analytic expressions, we refrain from such a description, which depends crucially on the material used.

\small
\bibliography{bibliography}

\begin{thebibliography}{33}%
\makeatletter
\providecommand \@ifxundefined [1]{%
 \@ifx{#1\undefined}
}%
\providecommand \@ifnum [1]{%
 \ifnum #1\expandafter \@firstoftwo
 \else \expandafter \@secondoftwo
 \fi
}%
\providecommand \@ifx [1]{%
 \ifx #1\expandafter \@firstoftwo
 \else \expandafter \@secondoftwo
 \fi
}%
\providecommand \natexlab [1]{#1}%
\providecommand \enquote  [1]{``#1''}%
\providecommand \bibnamefont  [1]{#1}%
\providecommand \bibfnamefont [1]{#1}%
\providecommand \citenamefont [1]{#1}%
\providecommand \href@noop [0]{\@secondoftwo}%
\providecommand \href [0]{\begingroup \@sanitize@url \@href}%
\providecommand \@href[1]{\@@startlink{#1}\@@href}%
\providecommand \@@href[1]{\endgroup#1\@@endlink}%
\providecommand \@sanitize@url [0]{\catcode `\\12\catcode `\$12\catcode
  `\&12\catcode `\#12\catcode `\^12\catcode `\_12\catcode `\%12\relax}%
\providecommand \@@startlink[1]{}%
\providecommand \@@endlink[0]{}%
\providecommand \url  [0]{\begingroup\@sanitize@url \@url }%
\providecommand \@url [1]{\endgroup\@href {#1}{\urlprefix }}%
\providecommand \urlprefix  [0]{URL }%
\providecommand \Eprint [0]{\href }%
\providecommand \doibase [0]{http://dx.doi.org/}%
\providecommand \selectlanguage [0]{\@gobble}%
\providecommand \bibinfo  [0]{\@secondoftwo}%
\providecommand \bibfield  [0]{\@secondoftwo}%
\providecommand \translation [1]{[#1]}%
\providecommand \BibitemOpen [0]{}%
\providecommand \bibitemStop [0]{}%
\providecommand \bibitemNoStop [0]{.\EOS\space}%
\providecommand \EOS [0]{\spacefactor3000\relax}%
\providecommand \BibitemShut  [1]{\csname bibitem#1\endcsname}%
\let\auto@bib@innerbib\@empty
%</preamble>
\bibitem [{\citenamefont {Einstein}\ \emph {et~al.}(1935)\citenamefont
  {Einstein}, \citenamefont {Podolsky},\ and\ \citenamefont
  {Rosen}}]{Einstein35}%
  \BibitemOpen
  \bibfield  {author} {\bibinfo {author} {\bibfnamefont {A.}~\bibnamefont
  {Einstein}}, \bibinfo {author} {\bibfnamefont {B.}~\bibnamefont {Podolsky}},
  \ and\ \bibinfo {author} {\bibfnamefont {N.}~\bibnamefont {Rosen}},\ }\href
  {\doibase 10.1103/PhysRev.47.777} {\bibfield  {journal} {\bibinfo  {journal}
  {Phys. Rev.}\ }\textbf {\bibinfo {volume} {47}},\ \bibinfo {pages} {777}
  (\bibinfo {year} {1935})}\BibitemShut {NoStop}%
\bibitem [{\citenamefont {Howell}\ \emph {et~al.}(2004)\citenamefont {Howell},
  \citenamefont {Bennink}, \citenamefont {Bentley},\ and\ \citenamefont
  {Boyd}}]{Howell04}%
  \BibitemOpen
  \bibfield  {author} {\bibinfo {author} {\bibfnamefont {J.~C.}\ \bibnamefont
  {Howell}}, \bibinfo {author} {\bibfnamefont {R.~S.}\ \bibnamefont {Bennink}},
  \bibinfo {author} {\bibfnamefont {S.~J.}\ \bibnamefont {Bentley}}, \ and\
  \bibinfo {author} {\bibfnamefont {R.~W.}\ \bibnamefont {Boyd}},\ }\href
  {\doibase 10.1103/PhysRevLett.92.210403} {\bibfield  {journal} {\bibinfo
  {journal} {Phys. Rev. Lett.}\ }\textbf {\bibinfo {volume} {92}},\ \bibinfo
  {pages} {210403} (\bibinfo {year} {2004})}\BibitemShut {NoStop}%
\bibitem [{\citenamefont {Law}\ and\ \citenamefont {Eberly}(2004)}]{Law04}%
  \BibitemOpen
  \bibfield  {author} {\bibinfo {author} {\bibfnamefont {C.~K.}\ \bibnamefont
  {Law}}\ and\ \bibinfo {author} {\bibfnamefont {J.~H.}\ \bibnamefont
  {Eberly}},\ }\href {\doibase 10.1103/PhysRevLett.92.127903} {\bibfield
  {journal} {\bibinfo  {journal} {Phys. Rev. Lett.}\ }\textbf {\bibinfo
  {volume} {92}},\ \bibinfo {pages} {127903} (\bibinfo {year}
  {2004})}\BibitemShut {NoStop}%
\bibitem [{\citenamefont {Monken}\ \emph {et~al.}(1998)\citenamefont {Monken},
  \citenamefont {Souto~Ribeiro},\ and\ \citenamefont {P\'adua}}]{Monken98}%
  \BibitemOpen
  \bibfield  {author} {\bibinfo {author} {\bibfnamefont {C.~H.}\ \bibnamefont
  {Monken}}, \bibinfo {author} {\bibfnamefont {P.~H.}\ \bibnamefont
  {Souto~Ribeiro}}, \ and\ \bibinfo {author} {\bibfnamefont {S.}~\bibnamefont
  {P\'adua}},\ }\href {\doibase 10.1103/PhysRevA.57.3123} {\bibfield  {journal}
  {\bibinfo  {journal} {Phys. Rev. A}\ }\textbf {\bibinfo {volume} {57}},\
  \bibinfo {pages} {3123} (\bibinfo {year} {1998})}\BibitemShut {NoStop}%
\bibitem [{\citenamefont {Chan}\ \emph {et~al.}(2007)\citenamefont {Chan},
  \citenamefont {Torres},\ and\ \citenamefont {Eberly}}]{Chan07}%
  \BibitemOpen
  \bibfield  {author} {\bibinfo {author} {\bibfnamefont {K.~W.}\ \bibnamefont
  {Chan}}, \bibinfo {author} {\bibfnamefont {J.~P.}\ \bibnamefont {Torres}}, \
  and\ \bibinfo {author} {\bibfnamefont {J.~H.}\ \bibnamefont {Eberly}},\
  }\href {\doibase 10.1103/PhysRevA.75.050101} {\bibfield  {journal} {\bibinfo
  {journal} {Phys. Rev. A}\ }\textbf {\bibinfo {volume} {75}},\ \bibinfo
  {pages} {050101} (\bibinfo {year} {2007})}\BibitemShut {NoStop}%
\bibitem [{\citenamefont {Gomes}\ \emph {et~al.}(2009)\citenamefont {Gomes},
  \citenamefont {Salles}, \citenamefont {Toscano}, \citenamefont
  {Souto~Ribeiro},\ and\ \citenamefont {Walborn}}]{Gomes09}%
  \BibitemOpen
  \bibfield  {author} {\bibinfo {author} {\bibfnamefont {R.~M.}\ \bibnamefont
  {Gomes}}, \bibinfo {author} {\bibfnamefont {A.}~\bibnamefont {Salles}},
  \bibinfo {author} {\bibfnamefont {F.}~\bibnamefont {Toscano}}, \bibinfo
  {author} {\bibfnamefont {P.~H.}\ \bibnamefont {Souto~Ribeiro}}, \ and\
  \bibinfo {author} {\bibfnamefont {S.~P.}\ \bibnamefont {Walborn}},\ }\href
  {\doibase 10.1103/PhysRevLett.103.033602} {\bibfield  {journal} {\bibinfo
  {journal} {Phys. Rev. Lett.}\ }\textbf {\bibinfo {volume} {103}},\ \bibinfo
  {pages} {033602} (\bibinfo {year} {2009})}\BibitemShut {NoStop}%
\bibitem [{\citenamefont {Menzel}\ \emph {et~al.}(2012)\citenamefont {Menzel},
  \citenamefont {Puhlmann}, \citenamefont {Heuer},\ and\ \citenamefont
  {Schleich}}]{Menzel12}%
  \BibitemOpen
  \bibfield  {author} {\bibinfo {author} {\bibfnamefont {R.}~\bibnamefont
  {Menzel}}, \bibinfo {author} {\bibfnamefont {D.}~\bibnamefont {Puhlmann}},
  \bibinfo {author} {\bibfnamefont {A.}~\bibnamefont {Heuer}}, \ and\ \bibinfo
  {author} {\bibfnamefont {W.~P.}\ \bibnamefont {Schleich}},\ }\href {\doibase
  10.1073/pnas.1201271109} {\bibfield  {journal} {\bibinfo  {journal} {Proc.
  Natl. Acad. Sci. U. S. A.}\ }\textbf {\bibinfo {volume} {109}},\ \bibinfo
  {pages} {9314} (\bibinfo {year} {2012})}\BibitemShut {NoStop}%
\bibitem [{\citenamefont {Menzel}\ \emph {et~al.}(2013)\citenamefont {Menzel},
  \citenamefont {Heuer}, \citenamefont {Puhlmann}, \citenamefont {Dechoum},
  \citenamefont {Hillery}, \citenamefont {Sp\"ahn},\ and\ \citenamefont
  {Schleich}}]{Menzel13}%
  \BibitemOpen
  \bibfield  {author} {\bibinfo {author} {\bibfnamefont {R.}~\bibnamefont
  {Menzel}}, \bibinfo {author} {\bibfnamefont {A.}~\bibnamefont {Heuer}},
  \bibinfo {author} {\bibfnamefont {D.}~\bibnamefont {Puhlmann}}, \bibinfo
  {author} {\bibfnamefont {K.}~\bibnamefont {Dechoum}}, \bibinfo {author}
  {\bibfnamefont {M.}~\bibnamefont {Hillery}}, \bibinfo {author} {\bibfnamefont
  {M.~J.~A.}\ \bibnamefont {Sp\"ahn}}, \ and\ \bibinfo {author} {\bibfnamefont
  {W.~P.}\ \bibnamefont {Schleich}},\ }\href {\doibase
  10.1080/09500340.2012.746400} {\bibfield  {journal} {\bibinfo  {journal} {J.
  Mod. Opt.}\ }\textbf {\bibinfo {volume} {60}},\ \bibinfo {pages} {86}
  (\bibinfo {year} {2013})}\BibitemShut {NoStop}%
\bibitem [{\citenamefont {Burlakov}\ \emph {et~al.}(2001)\citenamefont
  {Burlakov}, \citenamefont {Chekhova}, \citenamefont {Karabutova},\ and\
  \citenamefont {Kulik}}]{Burlakov01}%
  \BibitemOpen
  \bibfield  {author} {\bibinfo {author} {\bibfnamefont {A.~V.}\ \bibnamefont
  {Burlakov}}, \bibinfo {author} {\bibfnamefont {M.~V.}\ \bibnamefont
  {Chekhova}}, \bibinfo {author} {\bibfnamefont {O.~A.}\ \bibnamefont
  {Karabutova}}, \ and\ \bibinfo {author} {\bibfnamefont {S.~P.}\ \bibnamefont
  {Kulik}},\ }\href {\doibase 10.1103/PhysRevA.63.053801} {\bibfield  {journal}
  {\bibinfo  {journal} {Phys. Rev. A}\ }\textbf {\bibinfo {volume} {63}},\
  \bibinfo {pages} {053801} (\bibinfo {year} {2001})}\BibitemShut {NoStop}%
\bibitem [{\citenamefont {Jha}\ \emph {et~al.}(2008)\citenamefont {Jha},
  \citenamefont {O'Sullivan}, \citenamefont {Chan},\ and\ \citenamefont
  {Boyd}}]{Jha08}%
  \BibitemOpen
  \bibfield  {author} {\bibinfo {author} {\bibfnamefont {A.~K.}\ \bibnamefont
  {Jha}}, \bibinfo {author} {\bibfnamefont {M.~N.}\ \bibnamefont {O'Sullivan}},
  \bibinfo {author} {\bibfnamefont {K.~W.~C.}\ \bibnamefont {Chan}}, \ and\
  \bibinfo {author} {\bibfnamefont {R.~W.}\ \bibnamefont {Boyd}},\ }\href
  {\doibase 10.1103/PhysRevA.77.021801} {\bibfield  {journal} {\bibinfo
  {journal} {Phys. Rev. A}\ }\textbf {\bibinfo {volume} {77}},\ \bibinfo
  {pages} {021801} (\bibinfo {year} {2008})}\BibitemShut {NoStop}%
\bibitem [{\citenamefont {Kulkarni}\ \emph {et~al.}(2017)\citenamefont
  {Kulkarni}, \citenamefont {Kumar},\ and\ \citenamefont {Jha}}]{Kulkarni17}%
  \BibitemOpen
  \bibfield  {author} {\bibinfo {author} {\bibfnamefont {G.}~\bibnamefont
  {Kulkarni}}, \bibinfo {author} {\bibfnamefont {P.}~\bibnamefont {Kumar}}, \
  and\ \bibinfo {author} {\bibfnamefont {A.~K.}\ \bibnamefont {Jha}},\ }\href
  {\doibase 10.1364/JOSAB.34.001637} {\bibfield  {journal} {\bibinfo  {journal}
  {J. Opt. Soc. Am. B}\ }\textbf {\bibinfo {volume} {34}},\ \bibinfo {pages}
  {1637} (\bibinfo {year} {2017})}\BibitemShut {NoStop}%
\bibitem [{\citenamefont {Jha}\ and\ \citenamefont {Boyd}(2010)}]{Jha10}%
  \BibitemOpen
  \bibfield  {author} {\bibinfo {author} {\bibfnamefont {A.~K.}\ \bibnamefont
  {Jha}}\ and\ \bibinfo {author} {\bibfnamefont {R.~W.}\ \bibnamefont {Boyd}},\
  }\href {\doibase 10.1103/PhysRevA.81.013828} {\bibfield  {journal} {\bibinfo
  {journal} {Phys. Rev. A}\ }\textbf {\bibinfo {volume} {81}},\ \bibinfo
  {pages} {013828} (\bibinfo {year} {2010})}\BibitemShut {NoStop}%
\bibitem [{\citenamefont {Reid}(1989)}]{Reid89}%
  \BibitemOpen
  \bibfield  {author} {\bibinfo {author} {\bibfnamefont {M.~D.}\ \bibnamefont
  {Reid}},\ }\href {\doibase 10.1103/PhysRevA.40.913} {\bibfield  {journal}
  {\bibinfo  {journal} {Phys. Rev. A}\ }\textbf {\bibinfo {volume} {40}},\
  \bibinfo {pages} {913} (\bibinfo {year} {1989})}\BibitemShut {NoStop}%
\bibitem [{\citenamefont {Teh}\ and\ \citenamefont {Reid}(2014)}]{Teh14}%
  \BibitemOpen
  \bibfield  {author} {\bibinfo {author} {\bibfnamefont {R.~Y.}\ \bibnamefont
  {Teh}}\ and\ \bibinfo {author} {\bibfnamefont {M.~D.}\ \bibnamefont {Reid}},\
  }\href {\doibase 10.1103/PhysRevA.90.062337} {\bibfield  {journal} {\bibinfo
  {journal} {Phys. Rev. A}\ }\textbf {\bibinfo {volume} {90}},\ \bibinfo
  {pages} {062337} (\bibinfo {year} {2014})}\BibitemShut {NoStop}%
\bibitem [{\citenamefont {Hong}\ and\ \citenamefont {Mandel}(1985)}]{Hong85}%
  \BibitemOpen
  \bibfield  {author} {\bibinfo {author} {\bibfnamefont {C.~K.}\ \bibnamefont
  {Hong}}\ and\ \bibinfo {author} {\bibfnamefont {L.}~\bibnamefont {Mandel}},\
  }\href {\doibase 10.1103/PhysRevA.31.2409} {\bibfield  {journal} {\bibinfo
  {journal} {Phys. Rev. A}\ }\textbf {\bibinfo {volume} {31}},\ \bibinfo
  {pages} {2409} (\bibinfo {year} {1985})}\BibitemShut {NoStop}%
\bibitem [{\citenamefont {Freedman}\ and\ \citenamefont
  {Clauser}(1972)}]{Freedman72}%
  \BibitemOpen
  \bibfield  {author} {\bibinfo {author} {\bibfnamefont {S.~J.}\ \bibnamefont
  {Freedman}}\ and\ \bibinfo {author} {\bibfnamefont {J.~F.}\ \bibnamefont
  {Clauser}},\ }\href {\doibase 10.1103/PhysRevLett.28.938} {\bibfield
  {journal} {\bibinfo  {journal} {Phys. Rev. Lett.}\ }\textbf {\bibinfo
  {volume} {28}},\ \bibinfo {pages} {938} (\bibinfo {year} {1972})}\BibitemShut
  {NoStop}%
\bibitem [{\citenamefont {Mair}\ \emph {et~al.}(2001)\citenamefont {Mair},
  \citenamefont {Vaziri}, \citenamefont {Weihs},\ and\ \citenamefont
  {Zeilinger}}]{Mair01}%
  \BibitemOpen
  \bibfield  {author} {\bibinfo {author} {\bibfnamefont {A.}~\bibnamefont
  {Mair}}, \bibinfo {author} {\bibfnamefont {A.}~\bibnamefont {Vaziri}},
  \bibinfo {author} {\bibfnamefont {G.}~\bibnamefont {Weihs}}, \ and\ \bibinfo
  {author} {\bibfnamefont {A.}~\bibnamefont {Zeilinger}},\ }\href {\doibase
  10.1038/35085529} {\bibfield  {journal} {\bibinfo  {journal} {Nature}\
  }\textbf {\bibinfo {volume} {412}},\ \bibinfo {pages} {313} (\bibinfo {year}
  {2001})}\BibitemShut {NoStop}%
\bibitem [{\citenamefont {Leach}\ \emph {et~al.}(2010)\citenamefont {Leach},
  \citenamefont {Jack}, \citenamefont {Romero}, \citenamefont {Jha},
  \citenamefont {Yao}, \citenamefont {Franke-Arnold}, \citenamefont {Ireland},
  \citenamefont {Boyd}, \citenamefont {Barnett},\ and\ \citenamefont
  {Padgett}}]{Leach10}%
  \BibitemOpen
  \bibfield  {author} {\bibinfo {author} {\bibfnamefont {J.}~\bibnamefont
  {Leach}}, \bibinfo {author} {\bibfnamefont {B.}~\bibnamefont {Jack}},
  \bibinfo {author} {\bibfnamefont {J.}~\bibnamefont {Romero}}, \bibinfo
  {author} {\bibfnamefont {A.~K.}\ \bibnamefont {Jha}}, \bibinfo {author}
  {\bibfnamefont {A.~M.}\ \bibnamefont {Yao}}, \bibinfo {author} {\bibfnamefont
  {S.}~\bibnamefont {Franke-Arnold}}, \bibinfo {author} {\bibfnamefont {D.~G.}\
  \bibnamefont {Ireland}}, \bibinfo {author} {\bibfnamefont {R.~W.}\
  \bibnamefont {Boyd}}, \bibinfo {author} {\bibfnamefont {S.~M.}\ \bibnamefont
  {Barnett}}, \ and\ \bibinfo {author} {\bibfnamefont {M.~J.}\ \bibnamefont
  {Padgett}},\ }\href {\doibase 10.1126/science.1190523} {\bibfield  {journal}
  {\bibinfo  {journal} {Science}\ }\textbf {\bibinfo {volume} {329}},\ \bibinfo
  {pages} {662} (\bibinfo {year} {2010})}\BibitemShut {NoStop}%
\bibitem [{\citenamefont {Franson}(1989)}]{Franson89}%
  \BibitemOpen
  \bibfield  {author} {\bibinfo {author} {\bibfnamefont {J.~D.}\ \bibnamefont
  {Franson}},\ }\href {\doibase 10.1103/PhysRevLett.62.2205} {\bibfield
  {journal} {\bibinfo  {journal} {Phys. Rev. Lett.}\ }\textbf {\bibinfo
  {volume} {62}},\ \bibinfo {pages} {2205} (\bibinfo {year}
  {1989})}\BibitemShut {NoStop}%
\bibitem [{\citenamefont {Kwiat}\ \emph {et~al.}(1993)\citenamefont {Kwiat},
  \citenamefont {Steinberg},\ and\ \citenamefont {Chiao}}]{Kwiat93}%
  \BibitemOpen
  \bibfield  {author} {\bibinfo {author} {\bibfnamefont {P.~G.}\ \bibnamefont
  {Kwiat}}, \bibinfo {author} {\bibfnamefont {A.~M.}\ \bibnamefont
  {Steinberg}}, \ and\ \bibinfo {author} {\bibfnamefont {R.~Y.}\ \bibnamefont
  {Chiao}},\ }\href {\doibase 10.1103/PhysRevA.47.R2472} {\bibfield  {journal}
  {\bibinfo  {journal} {Phys. Rev. A}\ }\textbf {\bibinfo {volume} {47}},\
  \bibinfo {pages} {R2472} (\bibinfo {year} {1993})}\BibitemShut {NoStop}%
\bibitem [{\citenamefont {D'Angelo}\ \emph {et~al.}(2004)\citenamefont
  {D'Angelo}, \citenamefont {Kim}, \citenamefont {Kulik},\ and\ \citenamefont
  {Shih}}]{DAngelo04}%
  \BibitemOpen
  \bibfield  {author} {\bibinfo {author} {\bibfnamefont {M.}~\bibnamefont
  {D'Angelo}}, \bibinfo {author} {\bibfnamefont {Y.-H.}\ \bibnamefont {Kim}},
  \bibinfo {author} {\bibfnamefont {S.~P.}\ \bibnamefont {Kulik}}, \ and\
  \bibinfo {author} {\bibfnamefont {Y.}~\bibnamefont {Shih}},\ }\href {\doibase
  10.1103/PhysRevLett.92.233601} {\bibfield  {journal} {\bibinfo  {journal}
  {Phys. Rev. Lett.}\ }\textbf {\bibinfo {volume} {92}},\ \bibinfo {pages}
  {233601} (\bibinfo {year} {2004})}\BibitemShut {NoStop}%
\bibitem [{\citenamefont {Schneeloch}\ and\ \citenamefont
  {Howell}(2016)}]{Schneeloch16}%
  \BibitemOpen
  \bibfield  {author} {\bibinfo {author} {\bibfnamefont {J.}~\bibnamefont
  {Schneeloch}}\ and\ \bibinfo {author} {\bibfnamefont {J.~C.}\ \bibnamefont
  {Howell}},\ }\href {http://stacks.iop.org/2040-8986/18/i=5/a=053501}
  {\bibfield  {journal} {\bibinfo  {journal} {J. Opt. (Bristol, U. K.)}\
  }\textbf {\bibinfo {volume} {18}},\ \bibinfo {pages} {053501} (\bibinfo
  {year} {2016})}\BibitemShut {NoStop}%
\bibitem [{\citenamefont {Rubin}(1996)}]{Rubin96}%
  \BibitemOpen
  \bibfield  {author} {\bibinfo {author} {\bibfnamefont {M.~H.}\ \bibnamefont
  {Rubin}},\ }\href {\doibase 10.1103/PhysRevA.54.5349} {\bibfield  {journal}
  {\bibinfo  {journal} {Phys. Rev. A}\ }\textbf {\bibinfo {volume} {54}},\
  \bibinfo {pages} {5349} (\bibinfo {year} {1996})}\BibitemShut {NoStop}%
\bibitem [{\citenamefont {Marchand}\ and\ \citenamefont
  {Wolf}(1972)}]{Marchand72}%
  \BibitemOpen
  \bibfield  {author} {\bibinfo {author} {\bibfnamefont {E.~W.}\ \bibnamefont
  {Marchand}}\ and\ \bibinfo {author} {\bibfnamefont {E.}~\bibnamefont
  {Wolf}},\ }\href {\doibase 10.1364/JOSA.62.000379} {\bibfield  {journal}
  {\bibinfo  {journal} {J. Opt. Soc. Am.}\ }\textbf {\bibinfo {volume} {62}},\
  \bibinfo {pages} {379} (\bibinfo {year} {1972})}\BibitemShut {NoStop}%
\bibitem [{\citenamefont {Mandel}\ and\ \citenamefont {Wolf}(1995)}]{Mandel95}%
  \BibitemOpen
  \bibfield  {author} {\bibinfo {author} {\bibfnamefont {L.}~\bibnamefont
  {Mandel}}\ and\ \bibinfo {author} {\bibfnamefont {E.}~\bibnamefont {Wolf}},\
  }\href@noop {} {\emph {\bibinfo {title} {Optical coherence and quantum
  optics}}}\ (\bibinfo  {publisher} {Cambridge University Press},\ \bibinfo
  {address} {Cambridge},\ \bibinfo {year} {1995})\BibitemShut {NoStop}%
\bibitem [{\citenamefont {Schleich}(2001)}]{Schleich01}%
  \BibitemOpen
  \bibfield  {author} {\bibinfo {author} {\bibfnamefont {W.~P.}\ \bibnamefont
  {Schleich}},\ }\href@noop {} {\emph {\bibinfo {title} {Quantum optics in
  phase space}}}\ (\bibinfo  {publisher} {Wiley-VCH},\ \bibinfo {address}
  {Weinheim},\ \bibinfo {year} {2001})\BibitemShut {NoStop}%
\bibitem [{\citenamefont {Boyd}(2008)}]{Boyd08}%
  \BibitemOpen
  \bibfield  {author} {\bibinfo {author} {\bibfnamefont {R.~W.}\ \bibnamefont
  {Boyd}},\ }\href@noop {} {\emph {\bibinfo {title} {Nonlinear {O}ptics}}},\
  \bibinfo {edition} {3rd}\ ed.\ (\bibinfo  {publisher} {Academic Press},\
  \bibinfo {address} {Cambridge},\ \bibinfo {year} {2008})\BibitemShut
  {NoStop}%
\bibitem [{\citenamefont {Klyshko}(1993)}]{Klyshko93}%
  \BibitemOpen
  \bibfield  {author} {\bibinfo {author} {\bibfnamefont {D.~N.}\ \bibnamefont
  {Klyshko}},\ }\href {http://www.jetp.ac.ru/cgi-bin/dn/e_077_02_0222.pdf}
  {\bibfield  {journal} {\bibinfo  {journal} {J. Exp. Theor. Phys.}\ }\textbf
  {\bibinfo {volume} {104}},\ \bibinfo {pages} {2676} (\bibinfo {year}
  {1993})}\BibitemShut {NoStop}%
\bibitem [{\citenamefont {Walborn}\ \emph {et~al.}(2010)\citenamefont
  {Walborn}, \citenamefont {Monken}, \citenamefont {P\'adua},\ and\
  \citenamefont {Souto~Ribeiro}}]{Walborn10}%
  \BibitemOpen
  \bibfield  {author} {\bibinfo {author} {\bibfnamefont {S.~P.}\ \bibnamefont
  {Walborn}}, \bibinfo {author} {\bibfnamefont {C.~H.}\ \bibnamefont {Monken}},
  \bibinfo {author} {\bibfnamefont {S.}~\bibnamefont {P\'adua}}, \ and\
  \bibinfo {author} {\bibfnamefont {P.~H.}\ \bibnamefont {Souto~Ribeiro}},\
  }\href {\doibase https://doi.org/10.1016/j.physrep.2010.06.003} {\bibfield
  {journal} {\bibinfo  {journal} {Phys. Rep.}\ }\textbf {\bibinfo {volume}
  {495}},\ \bibinfo {pages} {87 } (\bibinfo {year} {2010})}\BibitemShut
  {NoStop}%
\bibitem [{\citenamefont {Di~Lorenzo~Pires}\ \emph {et~al.}(2011)\citenamefont
  {Di~Lorenzo~Pires}, \citenamefont {Coppens},\ and\ \citenamefont {van
  Exter}}]{DiLorenzoPires11}%
  \BibitemOpen
  \bibfield  {author} {\bibinfo {author} {\bibfnamefont {H.}~\bibnamefont
  {Di~Lorenzo~Pires}}, \bibinfo {author} {\bibfnamefont {F.~M. G.~J.}\
  \bibnamefont {Coppens}}, \ and\ \bibinfo {author} {\bibfnamefont {M.~P.}\
  \bibnamefont {van Exter}},\ }\href {\doibase 10.1103/PhysRevA.83.033837}
  {\bibfield  {journal} {\bibinfo  {journal} {Phys. Rev. A}\ }\textbf {\bibinfo
  {volume} {83}},\ \bibinfo {pages} {033837} (\bibinfo {year}
  {2011})}\BibitemShut {NoStop}%
\bibitem [{\citenamefont {Yun}\ \emph {et~al.}(2012)\citenamefont {Yun},
  \citenamefont {Xu}, \citenamefont {Zhao}, \citenamefont {Gong}, \citenamefont
  {Bai}, \citenamefont {Shi},\ and\ \citenamefont {Zhu}}]{Yun12}%
  \BibitemOpen
  \bibfield  {author} {\bibinfo {author} {\bibfnamefont {S.}~\bibnamefont
  {Yun}}, \bibinfo {author} {\bibfnamefont {P.}~\bibnamefont {Xu}}, \bibinfo
  {author} {\bibfnamefont {J.~S.}\ \bibnamefont {Zhao}}, \bibinfo {author}
  {\bibfnamefont {Y.~X.}\ \bibnamefont {Gong}}, \bibinfo {author}
  {\bibfnamefont {Y.~F.}\ \bibnamefont {Bai}}, \bibinfo {author} {\bibfnamefont
  {J.}~\bibnamefont {Shi}}, \ and\ \bibinfo {author} {\bibfnamefont {S.~N.}\
  \bibnamefont {Zhu}},\ }\href {\doibase 10.1103/PhysRevA.86.023852} {\bibfield
   {journal} {\bibinfo  {journal} {Phys. Rev. A}\ }\textbf {\bibinfo {volume}
  {86}},\ \bibinfo {pages} {023852} (\bibinfo {year} {2012})}\BibitemShut
  {NoStop}%
\bibitem [{\citenamefont {Zhong}\ \emph {et~al.}(2015)\citenamefont {Zhong},
  \citenamefont {Xu}, \citenamefont {Lu},\ and\ \citenamefont {Zhu}}]{Zhong15}%
  \BibitemOpen
  \bibfield  {author} {\bibinfo {author} {\bibfnamefont {M.}~\bibnamefont
  {Zhong}}, \bibinfo {author} {\bibfnamefont {P.}~\bibnamefont {Xu}}, \bibinfo
  {author} {\bibfnamefont {L.}~\bibnamefont {Lu}}, \ and\ \bibinfo {author}
  {\bibfnamefont {S.}~\bibnamefont {Zhu}},\ }\href {\doibase
  10.1364/JOSAB.32.002081} {\bibfield  {journal} {\bibinfo  {journal} {J. Opt.
  Soc. Am. B}\ }\textbf {\bibinfo {volume} {32}},\ \bibinfo {pages} {2081}
  (\bibinfo {year} {2015})}\BibitemShut {NoStop}%
\bibitem [{\citenamefont {Calder\'{o}n-Losada}\ \emph
  {et~al.}(2016)\citenamefont {Calder\'{o}n-Losada}, \citenamefont
  {Fl\'{o}rez}, \citenamefont {Villabona-Monsalve},\ and\ \citenamefont
  {Valencia}}]{CalderonLosada16}%
  \BibitemOpen
  \bibfield  {author} {\bibinfo {author} {\bibfnamefont {O.}~\bibnamefont
  {Calder\'{o}n-Losada}}, \bibinfo {author} {\bibfnamefont {J.}~\bibnamefont
  {Fl\'{o}rez}}, \bibinfo {author} {\bibfnamefont {J.~P.}\ \bibnamefont
  {Villabona-Monsalve}}, \ and\ \bibinfo {author} {\bibfnamefont
  {A.}~\bibnamefont {Valencia}},\ }\href {\doibase 10.1364/OL.41.001165}
  {\bibfield  {journal} {\bibinfo  {journal} {Opt. Lett.}\ }\textbf {\bibinfo
  {volume} {41}},\ \bibinfo {pages} {1165} (\bibinfo {year}
  {2016})}\BibitemShut {NoStop}%
\end{thebibliography}%

\end{document}